\documentclass[preprint,amsmath,amssymb,aps,a4]{revtex4-2}
\usepackage{graphicx}
\usepackage{bm}
\usepackage{txfonts}
\usepackage{hyperref}
\usepackage{mathrsfs}
\usepackage[english]{babel}
\usepackage{cleveref}
\usepackage[autostyle, english = american]{csquotes}
\MakeOuterQuote{"}
\date{\today}
\newcommand{\be}{\begin{eqnarray}}
	\newcommand{\ee}{\end{eqnarray}}

\newcommand{\bfz}{{\bf 0}_{\perp}}

\newcommand{\bfk}{{\bf k}_{\perp}}

\newcommand{\bfkj}{{\bf k}_{\perp j}}

\newcommand{\bfP}{{\bf P}_{\perp}}

\usepackage{xcolor}
\begin{document}
	
\title{Pions valence quark distributions in asymmetric nuclear matter at finite temperature}
	
	\author{Satyajit Puhan}
	\email{puhansatyajit@gmail.com}
	\affiliation{Department of Physics, Dr. B.R. Ambedkar National
		Institute of Technology, Jalandhar, 144008, India}
	
	\author{Navpreet Kaur}
	\email{knavpreet.hep@gmail.com}
	\affiliation{Department of Physics, Dr. B.R. Ambedkar National
		Institute of Technology, Jalandhar, 144008, India}
\author{Arvind Kumar}
\email{kumara@nitj.ac.in}
\affiliation{Department of Physics, Dr. B.R. Ambedkar National
	Institute of Technology, Jalandhar, 144008, India}
	
	\author{Suneel Dutt}
	\email{dutts@nitj.ac.in}
	\affiliation{Department of Physics, Dr. B.R. Ambedkar National
		Institute of Technology, Jalandhar, 144008, India}
	
		\author{Harleen Dahiya}
	\email{dahiyah@nitj.ac.in}
	\affiliation{Department of Physics, Dr. B.R. Ambedkar National
		Institute of Technology, Jalandhar, 144008, India}
\date{\today}%
%
%
\begin{abstract}

We have calculated the valence quark distributions of the lightest pseudoscalar meson, pions, in the isospin asymmetric nuclear matter at zero and finite temperature employing a light-cone quark model.
The medium modifications in the pion properties have been stimulated through the
effective  quark masses computed using the chiral SU($3$) quark mean field model. We have primarily focused  on the impact of isospin asymmetric medium  on the distribution amplitudes (DAs)  and parton distribution functions  (PDFs) of a valence quark 
for different baryon density and temperature values. Also, the DAs and PDFs have been evolved to $Q^2=10$ GeV$^2$ and $Q^2=16$ GeV$^2$ for different densities of nuclear medium and results have been compared with the available experimental data. The DAs and PDFs are found to modify substantially as a function of baryon density as compared to temperature and isospin asymmetry of the medium. 
\end{abstract}
\maketitle
\newpage
%
%
\section{Introduction\label{secintro}}

The study of  modifications in the properties of hadrons when subjected to dense nuclear environment is significant for understanding the experimental observables of high energy physics experiments and also in the exploration of quantum chromodynamics (QCD) properties such as chiral symmetry breaking at low energy and its restoration at higher densities and temperature.  The asymptotic freedom and color confinement are two properties of QCD realized at opposite sides of energy and momentum scale.  Due to color confinement, quarks and gluons are confined inside the hadrons and  their internal dynamics have rich contributions towards the vacuum properties of hadrons. However, finite density of nuclear medium may have considerable impact on the internal structure of hadrons as
has been indicated initially by the European Muon Collaboration   \cite{EuropeanMuon:1983wih}.  The nucleon structure functions in the nuclear medium  have been observed to be different as compared to that in the free space which is known as the EMC effect. Along with this, deeply bound pionic atom experiment \cite{Suzuki:2002ae} and pion-nucleus scattering have also represented the reduction of pion decay constant in the nuclear medium. Hence, one can naturally expect the modification of hadronic properties in nuclear medium which may be attributed to the partial restoration of chiral symmetry. The evidence of this has been authenticated through the low-energy pion-nucleus scattering \cite{Friedman:2004jh}, di-pion production in hadron-nucleus and  photon-nucleus processes \cite{CHAOS:1996nql,CHAOS:2004rhl}.  The importance of the study of in-medium properties of hadrons has also been realized through their role in understanding the dilepton spectra observed through the decay of vector mesons \cite{Fuchs:2004uu}. The medium modifications of electromagnetic properties of  hadrons using different theoretical approaches has also got considerable attention because of their role in ongoing and future  experiments such as J-PARC at KEK \cite{Kumano:2022cje}, NICA at JINR and electron-ion collider (EIC) at USA \cite{AbdulKhalek:2021gbh}.

In the low-$Q^2$ region, valence quark distribution functions have been used to access the non perturbative aspects of QCD which can be further used as inputs to evolve the distributions. This leads to perturbative QCD phenomena which are accessible experimentally in the  high-$Q^2$ region \cite{Hutauruk:2016sug}. Light pseudoscalar mesons result from the dynamical chiral symmetry breaking \cite{Nambu:1961tp}. Several models \cite{Puhan:2023ekt,Raha:2008ve,Nam:2007gf,Bijnens:2002hp,daSilva:2012gf} including lattice simulations \cite{Koponen:2017fvm,Hedditch:2007ex} have been applied to study the internal structure of the lightest pseudoscalar meson i.e. pion in vacuum. However, very limited progress has been made to explore the in-medium properties of pions in terms of their internal structure \cite{Suzuki:1995vr,deMelo:2014gea}. In Ref. \cite{Suzuki:1995vr}, the Nambu–Jona-Lasinio (NJL) model has been used to study the pion structure function in the nuclear medium. The medium modification of electromagnetic form factors, charge radii and weak decay constants has been studied by using the combined approach of  light-front pion wave function based on a Bethe-Salpeter amplitude model and quark meson coupling model (QMC) \cite{deMelo:2014gea}.  The quark meson coupling model has also been used in conjunction with NJL model to explore the electroweak properties of pions  in the nuclear medium \cite{Hutauruk:2018qku}.  The valence quark distributions of pions and kaons have been studied using NJL and QMC model  \cite{Hutauruk:2019ipp}.  The proper time regularization scheme has been implemented  \cite{Hutauruk:2021kej} to obtain the gluon and valence quark distributions of pions and kaons in the nuclear medium. However, in these works, the properties of pions have been investigated in the isospin symmetric nuclear matter at zero temperature only.

In the present work, we have investigated the distribution amplitudes (DAs) and  parton (valence quark) distribution functions
(PDFs) \cite{Martin:1998sq,Meissner:2008ay} of pions in the isospin asymmetric nuclear matter at zero and finite temperature. PDFs play an important role in describing the one dimensional internal structure of hadrons as a function of longitudinal momentum fraction ($x$). Similarly, for multi dimensional structure, one can study the transverse momentum-dependent parton distribution functions (TMDs) \cite{Diehl:2015uka,Angeles-Martinez:2015sea,Pasquini:2008ax}, generalized parton distributions (GPDs) \cite{Diehl:2003ny,Kaur:2023zhn}, generalized transverse momentum dependent parton distributions (GTMDs) \cite{Meissner:2008ay} etc. DAs give precise information about the strength of the coupling between quark and antiquark pairs within a hadron. It further quantifies the likelihood amplitude of discovering these constituent particles bound together, demonstrating the efficiency of their interaction in generating the hadronic structure \cite{Serna:2020txe,Chernyak:1983ej}. In our present calculations, the properties of pions  (DAs and valence quark distribution functions) have been calculated using the light-cone quark model (LCQM)  \cite{Brodsky:1997de,Qian:2008px,Kaur:2019jow}  whereas the in-medium effects have been stimulated using the chiral quark mean field (CQMF) model  \cite{Wang:2001jw}. LCQM is based on the light-cone formalism that provides a convenient framework to visualize the relativistic effects of partonic dynamics inside the hadron. It is a non-perturbative approach framework describing the structure and properties of hadrons and is is gauge invariant and relativistic by nature. LCQM focuses on
the valence quarks of the hadrons as they are the primary constituents responsible for the overall structure and properties of hadrons \cite{Puhan:2023ekt,Kaur:2020vkq}. In the low-$Q^2$ regime,  LCQM  has been successful in portraying the physical properties such as electromagnetic properties, charge radii, mass eigen values, decay constants, distribution amplitudes and  PDFs of pion (in the free space) with  comparable experimental results \cite{Kaur:2020vkq,Xiao:2002iv,Ma:1993ht}. The light-cone Fock state of a pseudoscalar pion in terms of its constitutent quark $q$, antiquark $\bar{q}$ and gluon $g$ can be expanded as $|\mathcal{P} \rangle =\sum |q \bar{q} \rangle \psi_{q\bar{q}} +\sum |q \bar{q} g \rangle \psi_{q\bar{q}g}+.... $. We have chosen to investigate the leading twist distributions by considering only the quark-antiquark Fock state.  The information about the bound-state dynamics as well as the momentum distribution between $\pi$ constituents can be scrutinized through DAs \cite{Brodsky:1989pv}.  DAs can be defined as an integral over the valence light-cone Fock state \cite{Lepage:1980fj}, whereas the PDFs present the tendency of a valence quark to carry a longitudinal momentum fraction $x$ \cite{Soper:1996sn,Martin:2009iq}.

Further, in order to explore the impact of isospin asymmetry and finite temperature of nuclear matter on the  DAs and PDFs of pions we have used the  CQMF model \cite{Wang:2001jw}. Compared to relativistic hadronic mean field models \cite{Papazoglou:1998vr}, where hadrons are treated as degrees of freedom,  the quarks are the fundamental building blocks   in the CQMF model confined inside the baryons through a confining potential. The constituent quarks confined inside the baryons interact through the exchange of  nonstrange scalar-isoscalar field $\sigma$, the strange scalar-isoscalar field $\zeta$ and the scalar-isovector field $\delta$.  Earlier, the CQMF model has been used to study the properties of nuclear \cite{Wang:2001jw} and strange hadronic matter \cite{Wang:2001hw}.  The model has also been used   to calculate the magnetic moments of octet and decuplet   baryons in dense nuclear \cite{Singh:2016hiw,Singh:2017mxj,Singh:2018kwq} and   strange matter \cite{Singh:2020nwp,Kumar:2023owb}. In light of the above developments, it is interesting to use the medium modified masses of constituent quarks calculated using CQMF model as inputs in the LCQM to obtain the modification of DAs and PDFs of pions in the asymmetric nuclear medium. 
  
The present work is organized as follows: In Sec. \ref{sec:chiral_mean_field} we describe the CQMF model used in the present work to obtain the in-medium constituent quark masses which have been used as inputs to obtain the  properties of pions in the nuclear matter.
The LCQM used to calculate the DAs and PDFs is presented in Sec. \ref{Sec:LCQM}.  In Sec. \ref{Sec:results}, the results of present work are discussed and finally summarized in Sec. \ref{Sec:conclusion}.

\section{Chiral SU(3) quark mean field model (CQMF) for isospin asymmetric nuclear matter} \label{sec:chiral_mean_field}
In the present section, we have detailed  the essentials of the CQMF model through which the medium modifications in 
the DAs and PDFs of pions are stimulated in LCQM. The constituent quarks in this model are bound inside the baryons through a confining potential and interact via the exchange of scalar fields $\sigma$ and $\zeta$. The scalar-isovector field $\delta$ contributes when the medium has finite isospin asymmetry. 
The low energy properties of QCD: chiral symmetry and its spontaneous breaking are incorporated in the model. Further, in the present CQMF model, the broken scale invariance property of QCD is also stimulated  through the dilaton field $\chi$.  
The general effective Lagrangian density of CQMF model is expressed as \cite{Wang:2001jw}
\begin{equation}
{\cal L}_{{\rm eff}} \, = \, {\cal L}_{q0} \, + \, {\cal L}_{qm}
\, + \,
{\cal L}_{\Sigma\Sigma} \,+\, {\cal L}_{VV} \,+\, {\cal L}_{\chi SB}\,  + \, {\cal L}_{c}. \label{totallag}
\end{equation}
The first term,  ${\cal L}_{q0} =\bar q \, i\gamma^\mu \partial_\mu \, q$ is the kinetic term for quarks. The second term, 
 ${\cal L}_{qm}$ of Eq. (\ref{totallag}) describes the interactions of constituent quarks with the scalar and vector mesons.
It is defined through relation  \cite{Wang:2001jw,Singh:2017mxj}
\begin{align}
{\cal L}_{qm}=g_s\left(\bar{\Psi}_LM\Psi_R+\bar{\Psi}_RM^{\dagger}\Psi_L\right)
-g_v\left(\bar{\Psi}_L\gamma^\mu l_\mu\Psi_L+\bar{\Psi}_R\gamma^\mu
r_\mu\Psi_R\right)~~~~~~~~~~~~~~~~~~~~~~~  \nonumber \\
=\frac{g_s}{\sqrt{2}}\bar{\Psi}\left(\sum_{a=0}^8 s_a\lambda_a
+ i \gamma^5 \sum_{a=0}^8 p_a\lambda_a
\right)\Psi -\frac{g_v}{2\sqrt{2}}
\bar{\Psi}\left( \gamma^\mu \sum_{a=0}^8
 v_\mu^a\lambda_a
- \gamma^\mu\gamma^5 \sum_{a=0}^8
a_\mu^a\lambda_a\right)\Psi, \label{quarkmesons}
\end{align}
where $\Psi = (u, d, s)$ denotes the quark field for corresponding flavors. Also,  
the parameters $g_s$ and $g_v$ are the couplings of quarks with scalar and vector mesons fields.  
 The self-interactions of scalar mesons $\sigma, \zeta$ and $\delta$ and the dilaton field $\chi$ are described by the third term ${\cal L}_{\Sigma\Sigma}$ of Eq. (\ref{totallag})
 and is given as
  \cite{Wang:2001jw,Singh:2017mxj} 
\begin{align}
{\cal L}_{\Sigma\Sigma} =& -\frac{1}{2} \, k_0\chi^2
\left(\sigma^2+\zeta^2+\delta^2\right)+k_1 \left(\sigma^2+\zeta^2+\delta^2\right)^2
+k_2\left(\frac{\sigma^4}{2} +\frac{\delta^4}{2}+3\sigma^2\delta^2+\zeta^4\right)\nonumber \\ 
&+k_3\chi\left(\sigma^2-\delta^2\right)\zeta 
 -k_4\chi^4-\frac14\chi^4 {\rm ln}\frac{\chi^4}{\chi_0^4} +
\frac{\xi}
3\chi^4 {\rm ln}\left(\left(\frac{\left(\sigma^2-\delta^2\right)\zeta}{\sigma_0^2\zeta_0}\right)\left(\frac{\chi^3}{\chi_0^3}\right)\right). \label{scalar0}
\end{align} 
The last two logarithmic terms in this equation take into account the scale breaking effects and  help to express the trace of energy momentum tensor proportional to the fourth power of the dilaton field $\chi$ within the CQMF model \cite{Wang:2001jw,Papazoglou:1998vr}.
  The fourth term of Eq. (\ref{totallag}) gives the self interactions of vector mesons $\omega$ and $\rho$  which is expressed as
\begin{equation}
{\cal L}_{VV}=\frac{1}{2} \, \frac{\chi^2}{\chi_0^2} \left(
m_\omega^2\omega^2+m_\rho^2\rho^2\right)+g_4\left(\omega^4+6\omega^2\rho^2+\rho^4\right). 
\label{vector}
\end{equation}
In similar lines with the scalar-isovector field $\delta$, the vector-isovector field $\rho$ also contribute when the medium has finite isospin asymmetry. In Eq. (\ref{totallag}), the  term ${\cal L}_{\chi SB}$  representing the explicit symmetry breaking term is given by
\begin{equation}\label{L_SB}
{\cal L}_{\chi SB}=\frac{\chi^2}{\chi_0^2}\left[m_\pi^2\kappa_\pi\sigma +
\left(
\sqrt{2} \, m_K^2\kappa_K-\frac{m_\pi^2}{\sqrt{2}} \kappa_\pi\right)\zeta\right].
\end{equation}
The last term of Eq. (\ref{totallag}),
corresponds to the confinement of quarks inside the baryons and is defined as
\begin{align}
{\cal L}_{c} = -  \bar \Psi \chi_c \Psi.
\end{align}
Here, $ 
\chi_{c}(r)=\frac14 k_{c} \, r^2(1+\gamma^0)$ is the confining potential, where
the coupling constant $k_c$ is taken to be $98 \, \text{MeV}. \text{fm}^{-2}$ \cite{Wang:2001jw,Kumar:2023owb}.

To investigate the properties of isospin asymmetric nuclear matter at finite temperature and density we have used the mean field approximation. The Dirac equation, under the influence of meson mean field, for the quark field $\Psi_{qi}$, is given as 
\cite{Wang:2001jw}
\begin{equation}
\left[-i\vec{\alpha}\cdot\vec{\nabla}+\chi_c(r)+\beta m_q^*\right]
\Psi_{qi}=e_q^*\Psi_{qi}. \label{Dirac}
\end{equation}
Here, subscripts $q$ and $i$ denote the quark $q$ ($q=u, d, s$)
in a baryon of type $i$ ($i=p, n$). 
The effective quark mass $m_{q}^*$ is defined in terms of scalar fields $\sigma, \zeta$ and $\delta$ through relation
\begin{equation}
m_q^*=-g_\sigma^q\sigma - g_\zeta^q\zeta - g_\delta^q I^{3q} \delta + \Delta m, \label{qmass}
\end{equation}
where $\Delta m$ is zero for non-strange $u$ and $d$ quarks, whereas for strange $s$ quark $\Delta m=77$ MeV.
The term $\Delta m$ is introduced through the Lagrangian density   ${\cal L}_{\Delta m} = - (\Delta m) \bar \psi S_1 \psi$  to obtain a reasonable value for the vacuum mass of $s$ quark ($m_s = 450$ MeV) \cite{Wang:2001jw}. Here,  $S_1 \, = \, \frac{1}{3} \, \left(I - \lambda_8\sqrt{3}\right) = {\rm diag}(0,0,1)$ is the strange quark matrix.
 Effective energy of particular quark under the influence of meson field is given as
 \begin{equation} 
e_q^*=e_q-g_\omega^q\omega-g_\rho^q I^{3q}\rho\,.
\label{eq_eff_energy1}
 \end{equation}

Within the CQMF model, the effective mass $M_i^{*}$ of the $i^{th}$ baryon is related to the effective energy $E_i^*$ and in-medium spurious center of mass momentum $p_{i \, \text{cm}}^{*}$ \cite{Barik:1985rm,Barik:2013lna} through relation
\begin{align}
M_i^*=\sqrt{E_i^{*2}- \langle p_{i \, \text{cm}}^{*2} \rangle}\,. \label{baryonmass}
\end{align} 
The effective energy, $E_i^{*}$ can also be expressed in terms of effective energy $e_q^*$ of constituent quarks using relation 
\begin{equation} 
E_i^*=\sum_qn_{qi}e_q^*+E_{i \, \text{spin}}\,.
    \label{energy}
\end{equation}
In the above equation, 
$n_{qi}$ denotes the number of quarks of type $q$ present in the $i^{th}$ baryon.
Further, the term $E_{i \, \text{spin}}$  which
contributes as a correction to baryon energy due to spin-spin interaction is fitted to the vacuum masses of baryons.
The 
spurious center of mass momentum $p_{i \, \text{cm}}^{*}$ is related to the in-medium constituent quark mass $m_q^{*}$ and the effective energy $e_q^{*}$ through the relation
\cite{Barik:1985rm,Barik:2013lna}
\begin{equation}
\left\langle p_{i \mathrm{~cm}}^{* 2}\right\rangle=\frac{\left(11 e_q^*+m_q^*\right)}{6\left(3 e_q^*+m_q^*\right)}\left(e_q^{* 2}-m_q^{* 2}\right).
\end{equation}

The thermodynamic potential for the isospin asymmetric nuclear matter at finite temperature is written as 
\begin{equation}
\Omega = -\frac{k_{B}T}
{(2\pi)^3} \sum_{i} \gamma_i
\int_0^\infty d^3k\biggl\{{\rm ln}
\left( 1+e^{- [ E^{\ast}_i(k) - \nu_i^* ]/k_{B}T}\right) \\
+ {\rm ln}\left( 1+e^{- [ E^{\ast}_i(k)+\nu_i^* ]/k_{B}T}
\right) \biggr\} -{\cal L}_{M}-{\cal V}_{\text{vac}}, 
\label{Eq_therm_pot1}  
\end{equation}
where summation is over the nucleons of the medium, i.e., $i = p,n$ and
$\gamma_i=2$ is the degeneracy factor.
Also, $
{\cal L}_{M} \, = 
{\cal L}_{\Sigma\Sigma} \,+\, {\cal L}_{VV} \,+\, {\cal L}_{\chi SB}\,
$,  
and $E^{\ast }(k)=\sqrt{M_i^{\ast 2}+k^{2}}$.
The term ${\cal V}_{\text{vac}}$ is subtracted to obtain zero vacuum energy. The effective chemical potential $\nu_i^*$ of the nucleons is defined in terms of free chemical potential $\nu_i$ and is given by \cite{Wang:2001jw}
\begin{align}
\nu_i^* = \nu_i - g_{\omega}^i\omega -g_{\rho}^i I^{3i} \rho.
\end{align}
The thermodynamic potential defined in Eq. (\ref{Eq_therm_pot1}) is minimized with respect to the scalar fields
 $\sigma$, $\zeta$ and $\delta$,
 the dilaton field, $\chi$,  and, the vector fields $\omega$ and $\rho$ 
 through
 \begin{align}
  \frac{\partial \Omega}{\partial \sigma} = 
  \frac{\partial \Omega}{\partial \zeta} =
  \frac{\partial \Omega}{\partial \delta} =
  \frac{\partial \Omega}{\partial \chi} =
  \frac{\partial \Omega}{\partial \omega} =
  \frac{\partial \Omega}{\partial \rho}  =
    0.
    \label{eq:therm_min1}
  \end{align}
The system of non-linear equations  obtained for the scalar and vector fields are solved for finite baryon density, temperature and isospin asymmetry of the medium.   The isospin asymmetry of the nuclear medium is defined in terms of isospin asymmetry parameter,  $\eta = \frac{\rho_n - \rho_p}{2\rho_{B}}$.    Here, $\rho_B = \rho_p + \rho_n$ is the total baryonic density of the nuclear medium.    The finite temperature effects are introduced through the  temperature dependence of vector density, $\rho_i$ and the scalar density, $\rho_i^s$ of the nucleons respectively through the relations
    \begin{align}
    \rho_{i} = \gamma_{i} \int\frac{d^{3}k}{(2\pi)^{3}}  
    \Big(n_i(k)-\bar{n}_i(k)
    \Big),
    \label{rhov0}
    \end{align}
     and
    \begin{eqnarray}
    \rho_{i}^{s} = \gamma_{i} \int\frac{d^{3}k}{(2\pi)^{3}} 
    \frac{m_{i}^{*}}{E^{\ast}_i(k)} \Big(n_i(k)+\bar{n}_i(k)
    \Big)\,.
    \label{rhos0}
    \end{eqnarray}
Note that the scalar densities, $\rho_i^s$ appear as the source terms in the equations of motion for scalar fields $\sigma, \zeta$ and $\delta$, whereas, the number densities,  $\rho_i$ in the equations of motion of vector fields $\omega$ and $\rho$. In the above equations,   $n_i(k)$ and $\bar{n}_i(k)$ represent the Fermi distribution functions at finite temperature for fermions and anti-fermions and are respectively given as
   
\begin{equation}
    n_i(k) = \frac{1}{1+\exp\left[(E^*_{i}(k) 
    -\nu^{*}_{i})/k_BT \right]}~~~ \text{and}~~~
    \bar{n}_i(k) = \frac{1}{1+\exp\left[( E^*_{i}(k) 
    +\nu^{*}_{i})/k_BT\right]}~.
    \label{dfp}
\end{equation}

\section{Light-cone quark  model}
\label{Sec:LCQM}
For a hadron carrying total momentum $P$ with light-cone coordinates $(P^+,P^-,\bfP)$ and longitudinal spin projection $S_z$ in LCQM, the expansion of hadron eigenstate $|\mathcal{M}(P^+,\bfP,S_z)\rangle$ in terms of multiparticle Fock eigenstates $|n\rangle$ is expressed as \cite{Qian:2008px,Lepage:1980fj}
\be
|\mathcal{M}(P^+,\bfP,S_z)\rangle &=& \sum_{n,\lambda_j} \int \prod_{j=1}^{n} \frac{dx_j~  d^2\bfkj}{2(2\pi)^3\sqrt{x_{j}}} \, 16 \pi^{3} \, \delta \bigg(1-\sum_{j=1}^{n} x_{j}\bigg) \, \delta^{(2)} \bigg(\sum_{j=1}^{n}\bfkj\bigg) \nonumber \\		
&\times& \psi_{n/\mathcal{M}}(x_{j},\bfkj,\lambda_{j})|n; x_{j} P^{+},x_{j}\bfP + \bfkj,\lambda_{j}\rangle \, ,
\label{MesonState}\ee
where $x_j=\frac{\textbf{k}_j^+}{P^+}$, $\bfkj$ and $\lambda_j$ respectively correspond to the longitudinal momentum fraction, transverse momentum and helicity carried by the $j$th constituent parton. 
The multiparticle state of $n$-particles is normalized as 
\be
\langle n; k^{\prime +}_j, \bfkj^\prime, \lambda_{j}^\prime|n ; k^+_j, \bfkj, \lambda_j \rangle = \prod_{j=1}^{n} 16 \pi^{3} \,  k^{\prime +}_j \, \delta (k^{\prime +}_j-k^+_j) \, \delta^{(2)} ( \bfkj^\prime-\bfkj) \, \delta_{\lambda_{j}^\prime \lambda_{j}} \, .
\ee
In light-cone frame, the momenta of meson and its constituent quarks $u$($\bar{d}$), having effective  masses $M^*$ and $m_u^*$($m_{\bar{d}}^*$) respectively, are expressed as
\be 
P &=& \bigg(P^+,\frac{M^{\ast2}}{P^+},\bfz\bigg) \, ,  \nonumber \\
k_1 &=& \bigg(x_1 P^+,\frac{\bfk^2 + m^{\ast2}_u}{x_1 P^+},\bfk \bigg) \, ,  \nonumber \\
k_2 &=& \bigg(x_2 P^+,\frac{\bfk^2 + m^{\ast2}_{\bar{d}}}{x_2 P^+},-\bfk \bigg) \, . 
\ee
In accordance with constraint $\sum_{j=1}^{n} x_j = 1$ for the light-cone quark momentum in light-cone dynamics, we have $x_1 + x_2 =1$ implying that if quark carries $x$ fraction of longitudinal momentum, then anti-quark will be carrying the fraction $(1-x)$.
The expansion of two-particle Fock state in terms of light-cone wave functions (LCWFs) for  pseudoscalar $\pi$ meson having spin $S_z=0$  can be written as
\be 
|\mathcal{P} (P^+,\bfP,S_z)\rangle &=& \int \frac{dx \, d^2 \bfk}{  16 \pi^3 \sqrt{x(1-x)}} \, \big[\psi (x,\bfk,\uparrow,\uparrow) \, |x P^+, \bfk, \uparrow, \uparrow \rangle   \nonumber \\
&+& \psi (x,\bfk,\uparrow,\downarrow) \, |x P^+, \bfk, \uparrow, \downarrow \rangle + \psi (x,\bfk,\downarrow,\uparrow) \, |x P^+, \bfk,  \downarrow,\uparrow \rangle \nonumber \\ &+& \psi (x,\bfk,\downarrow,\downarrow) \, |x P^+, \bfk, \downarrow, \downarrow \rangle \big] \, \label{eqnq} .
\ee 
These LCWFs can be further expressed in terms of momentum space $\varphi$ and spin $\Phi$ wave functions as \cite{Huang:1994dy} 
\be 
\psi(x,\bfk,\lambda_1, \lambda_2)= \varphi(x,\bfk) \, \Phi (x,\bfk,\lambda_1, \lambda_2) \, ,
\ee 
where $\lambda_1(\lambda_2)$ represent the helicity of the quark (anti-quark). 
By making the use of Melosh-Wigner rotation \cite{Melosh:1974cu}, the transformed instant-form SU($6$) coefficient $\Phi(x,\bfk,\lambda_1, \lambda_2)$, in terms of light-cone components for $\pi$, can be expressed as \cite{Xiao:2002iv,Xiao:2003wf}
\be 
\Phi(x,\bfk,\uparrow,\downarrow)=\frac{1}{\sqrt{2}\varUpsilon_1 \varUpsilon_2} [(M^\ast x +m^\ast_u) (M^\ast (1-x)+m^\ast_{\bar{d}})-\bfk^2] \, , \nonumber \\
\Phi(x,\bfk,\downarrow,\uparrow)=-\frac{1}{\sqrt{2}\varUpsilon_1 \varUpsilon_2} [(M^\ast x +m^\ast_u) (M^\ast (1-x) +m^\ast_{\bar{d}})-\bfk^2] \, , \nonumber \\
\Phi(x,\bfk,\uparrow,\uparrow)=\frac{1}{\sqrt{2}\varUpsilon_1 \varUpsilon_2} [(M^\ast x +m^\ast_u) k_2^l - (M^{\ast} (1-x) +m^\ast_{\bar{d}}) k_1^l] \, , \nonumber \\
\Phi(x,\bfk,\downarrow,\downarrow)=\frac{1}{\sqrt{2}\varUpsilon_1 \varUpsilon_2} [(M^\ast x +m^\ast_u) k_2^r - (M^\ast (1-x) +m^\ast_{\bar{d}}) k_1^r] \, , 
\label{SpinWfns}
\ee  
where $\varUpsilon_1 = \sqrt{(M^\ast x +m_u^{\ast2})^2 + \bfk^2}$, $\varUpsilon_2 = \sqrt{(M^\ast (1-x) +m_{\bar{d}}^{\ast2})^2 + \bfk^2}$ and $k_{1(2)}^{r,l} = k_{1(2)}^1\pm k_{1(2)}^2$. The quantity $M^\ast$ satisfies the condition 
\be 
M^{\ast2} = \frac{m_u^{\ast2} + \bfk^2}{x} + \frac{m_{\bar{d}}^{\ast2} + \bfk^2}{1-x} \, .
\ee 
 The spin wave function given in Eq. (\ref{SpinWfns}) must satisfy the normalization condition
\be 
\sum_{\lambda_1 \lambda_2} \Phi^\ast (x,\bfk,\lambda_1,\lambda_2) \, \Phi(x,\bfk,\lambda_1,\lambda_2) = 1 \, .
\ee
For momentum space wave function, we have adopted the Brodsky-Huang-Lepage prescription \cite{Yu:2007hp,Xiao:2002iv,Kaur:2020vkq} which is represented as follows
\be 
\varphi (x,\bfk)=\mathcal{A} \, exp \,\Biggl[-\frac{ \frac{m_{u}^{\ast2} + \bfk^2}{x} + \frac{m_{\bar{d}}^{\ast2} + \bfk^2}{1-x}}{8 \beta^2} - \frac{(m_u^{\ast2} - m_{\bar{d}}^{\ast2})^2}{8 \beta^2 \, \bigg( \frac{m_u^{\ast2} + \bfk^2}{x} + \frac{m_{\bar{d}}^{\ast2} + \bfk^2}{1-x}\bigg)}\Biggr] \, , \label{momspace}
\ee 
where $\mathcal{A}= A \, exp \, \big[\frac{m_u^{\ast2} + m_{\bar{d}}^{\ast2}}{8 \beta^2}\big]$ with $A$ and $\beta$ are the normalization constant and harmonic scale parameter respectively. The momentum space wave function is normalized as
\begin{eqnarray}
    \int \frac{{d x} d^2 \bfk}{2 (2 \pi)^3} \, |\varphi (x,\bfk)|^2 =1 \, .
\end{eqnarray}

For the calculations of pion properties in asymmetric nuclear matter, the longitudinal momentum fraction of $j$th parton in the free space, i.e., $x_j = \frac{k_j^+}{P^+}$ has been replaced with their in-medium values, $x_j^* = \frac{k_j^{*+}}{P^{*+}}$.
Note that in the free space
\begin{align}
x_j = \frac{k_j^+}{P^+} = 
\frac{k_j^0 + k_j^3}{P^0+P^3},
\label{Eq_xfrac_free}
\end{align}
where $k_j^0 = E_j$ and $P^0 = E_\pi$ are the energies of $j$th quark and $\pi$ meson in the free space.
For the in-medium case,
Eq. (\ref{Eq_xfrac_free}) changes to 
\cite{Arifi:2023jfe}
\begin{align}
x_j^* = \frac{k_j^{*+}}{P^{*+}} = 
\frac{k_j^{*0} + k_j^{*3}}{P^{*0}+P^{*3}}.
\label{Eq_xfrac_med1}
\end{align}
Within the CQMF model, the in-medium values of $k_j^{*0}$
and $P^{*0}$
will have the contributions
from vector fields $\omega$ and $\rho$. For the $k_j^{*0}$ (quark case), we have
\begin{align}
k_j^{*0} = E_j^{*} + g_{\omega}^{j}\omega + 
g_{\rho}^{j} I^{3j}\rho,
\label{Eq_kj1}
\end{align}
where $E_{j}^{*} = \sqrt{k_j^2 +m_{j}^{*2}}$. For the antiquark, the above equation will modify to
\begin{align}
k_{\bar j}^{*0} = E_{\bar j}^{*} - g_{\omega}^{{\bar j}}\omega - 
g_{\rho}^{{\bar j}} I^{3{\bar j}}\rho.
\label{Eq_kjbar1}
\end{align}
The coupling strengths of quark and antiquark 
 with $\omega$ and $\rho$ fields are assumed to be same in the present work, i.e., $g_{\omega}^{{j}} = g_{\omega}^{{\bar j}}$ and
 $g_{\rho}^{{j}} = g_{\rho}^{{\bar j}}$.
 For the in-medium energy $P^{*0}$ of the $\pi$ meson (consisting of one light quark and one antiquark), we
 have
 \begin{align}
P^{*0} = k_{ j}^{*0} + k_{\bar j}^{*0} = E_j^* + E_{\bar j}^* + g_{\rho}^{{ j}} \left(I^{3{j}}-I^{3{\bar j}}\right)\rho.
\label{Eq_medP02}
 \end{align}
 Therefore, substituting Eqs. (\ref{Eq_kj1}), (\ref{Eq_kjbar1})  and (\ref{Eq_medP02}) in Eq. (\ref{Eq_xfrac_med1}),  the longitudinal momentum fraction for the $j$th quark/antiquark can be expressed as
 \begin{align}
 x_j^*  = 
 \begin{cases}
 \frac{E_j^* + g_{\omega}^{j}\omega + 
 g_{\rho}^{j} I^{3j}\rho + k_j^{*3}}{E_j^* + E_{\bar j}^* + g_{\rho}^{{ j}} \left(I^{3{j}}-I^{3{\bar j}}\right)\rho + P^{*3}} = \frac{x_j+ (g_{\omega}^{j}\omega + 
 g_{\rho}^{j} I^{3j}\rho)/P^+}{1+\left(I^{3{j}}-I^{3{\bar j}}\right)\rho/P^+} \quad \text{for quark } q \\
  \frac{E_{\bar j}^* - g_{\omega}^{j}\omega - 
  g_{\rho}^{j} I^{3\bar j}\rho + k_{\bar j}^{*3}}{E_j^* + E_{\bar j}^* + g_{\rho}^{{ j}} \left(I^{3{j}}-I^{3{\bar j}}\right)\rho + P^{*3}} = \frac{x_j - (g_{\omega}^{j}\omega + 
 g_{\rho}^{j} I^{3j}\rho)/P^+}{1+\left(I^{3{j}}-I^{3{\bar j}}\right)\rho/P^+} \quad
  \text{for antiquark } \bar{q} .
 \end{cases}
 \label{Eq_xfrac_med2}
 \end{align}
In Ref. \cite{Hutauruk:2019ipp}, at zero temperature in symmetric nuclear matter, the in-medium longitudinal momentum fraction $x$ has been expressed in terms of fermi energy $\epsilon_F$ of quark through relation
\be
x^*=\frac{\epsilon_F}{E_F}x -\frac{V^0_i}{E_F} \, ,
\ee
where $\epsilon_F=\sqrt{(k^{q }_F)^2+(m_q^*)^2} + V^0_i= E_F+V^0_i$ is for quark and $V^0_i$ refers to the vector potential which is expressed as $V^0=g^q_\omega \omega$. In our case, it can be interpreted as $V^0=g^q_\omega \omega+g_{\rho}^q I^{3i} \rho$). Fermi momentum of quark $k_F$ is related to the baryonic density as $\rho_B=2 (k_F)^3 / 3 \pi^2$. However, in the present work,  we have constrained our calculations to  only $x$ for simplification. 

\section{Results and discussion}
\label{Sec:results}
 
\subsection{In-Medium Effective Quark Masses}
In the present section, we have discussed the medium modifications on DAs and PDFs of pions in the asymmetric nuclear medium at finite temperature. 
The DAs and PDFs of pions calculated using the LCQM model have been modified in the nuclear medium through the
in-medium quark masses, $m_q^{*}$ calculated using Eq. (\ref{qmass}) within the CQMF model. The density and temperature dependent values of scalar fields $\sigma, \zeta$ and $\delta$ have been obtained by solving the coupled system of equation from  Eq.(\ref{eq:therm_min1}). 
The various parameters used as inputs to solve the system of equations of scalar and vector field within the CQMF model have been listed in Table (\ref{table1_para}). The   parameters $k_0, k_1, k_2, k_3$ and $k_4$ used in the Eq. (\ref{scalar0}) have been calculated using $\pi$ meson mass, $m_{\pi}$, $K$ meson mass, $m_K$ and the average mass of $\eta$ and $\eta^{'}$ mesons \cite{Wang:2001jw}. 
The vacuum expectation values of scalar meson fields $\sigma$ and $\zeta$ ($\sigma_0$ and $\zeta_0$) are related to the pion decay constant, $\kappa_\pi$ and the kaon decay constant, $\kappa_K$, through relations $\sigma_0= -\kappa_{\pi} ~~{\rm and}~~~~  \zeta_0= \frac{1}{\sqrt{2}}\left( \kappa_{\pi}-2\kappa_{K}\right)$ respectively. The vacuum value of dilaton field $\chi_0=254.6$ MeV and the coupling constant $g_4=37.4$ have been fitted to reasonable effective nucleon mass \cite{Wang:2001jw}. The value of parameter  $\xi$ appearing in the scale breaking terms of Eq. (\ref{scalar0}) has been determined from   the  QCD $\beta$-function at one loop level for three colors and three flavors \cite{Papazoglou:1998vr,Wang:2001jw}.
   
The coupling constants $g_\sigma^q$ and $g_\omega^N$ are fitted to obtain the binding energy of symmetric nuclear matter as $-16$ MeV at nuclear saturation density $\rho_{B}=0.16$ fm$^{-1}$. The vacuum constituent quark masses at $\rho_B=0$ (constituent quark mass of light $u/d$ quark in the free space) are found to be $m_u^{0} =m_d^{0} =256$ MeV. The coupling strength of non-strange scalar field $\sigma$ with the strange quark, $s$ and the  strange scalar field $\zeta$ with the light quarks, $u$ and $d$ is zero, i.e.,  $g_\sigma^s = g_\zeta^u = g_\zeta^d = 0 $. Also, for the scalar iso-vector field $ \delta$, the relations considered in the calculations  $g_{\delta}^u = g_{\sigma}^u$ and  $g_{\delta}^s = 0$.  For the vector mesons, various coupling strengths are related through relations $\frac{g_v}{2\sqrt{2}}   = g_{\rho}^u = g_{\rho}^d = g_\omega^u = g_\omega^d = \frac{1}{3} g_\omega^N$ and $g_\omega^s =   g_{\rho}^s  = 0$.
   
\begin{table}
	\begin{tabular}{|c|c|c|c|c|c|c|c|c|c|} 
		\hline 
		$k_0$ & $k_1$ & $k_2$ & $k_3$ & $k_4$ & $\sigma_0$ (MeV) & $\zeta_0$ (MeV) & $\chi_0$ (MeV) & $\xi$ & $\rho_0$ ($\text{fm}^{-3}$) \\ 
		\hline 
		$~$ 4.94 $~$ & $~$ 2.12 $~$ & $~$ $-$10.16 $~$ & $~$ $-$5.38 $~$ & $~$ $-$0.06 $~$ & $-$92.8 & $-$96.5 & 254.6 & $~$ 6/33 $~$ & 0.16  \\ 
		\hline
	\end{tabular}

 \begin{tabular}{|c|c|c|c|c|c|c|c|c|c|} 
		\hline 
	$g_{\sigma}^u=g_{\sigma}^d$ & $g_{\sigma}^s = g_{\zeta}^u=g_{\zeta}^d$ & $g_{\delta}^u$ & $g_{\zeta}^s = g_s$ & $g_4$ & $g_{\delta}^p = g_{\delta}^u$  & $g_{\omega}^N = 3g_{\omega }^u$  & $g_{\rho}^p$  & $m_{\pi}$ (MeV) & $m_K$ (MeV) \\ 
		\hline 
		2.72 & 0 & $~$ 2.72 $~$ & 3.847 & $~$ 37.4 $~$ & 2.72 & 9.69 & $~$ 8.886 $~$ & 139 & 494   \\ 
		\hline
	\end{tabular}
	\caption{Values of various input parameters used in the present work \cite{Wang:2001jw}.} \label{table1_para}
\end{table}

\begin{table}[h]
	\centering
	\begin{tabular}{|c|c|c|c|c|c|c|c|c|}
		\hline
		$~  ~$ & \multicolumn{8}{c|}{Effective quark masses (GeV)} \\
		\cline{2-9}
		$~$ Baryon density $~$ & \multicolumn{4}{c|}{ $T=0$ GeV}  & \multicolumn{4}{c|} {$T=0.1$ GeV}   \\
		\cline{2-9}
		$~ \rho_B/\rho_0 ~$ & \multicolumn{2}{c|}{$\eta=0$}  & \multicolumn{2}{c|}{$\eta=0.5$} & \multicolumn{2}{c|}{$\eta=0$}  & \multicolumn{2}{c|}{$\eta=0.5$}  \\
		\cline{2-9}
		$~$  $~$ & $~$ $m_u^*$  $~$ & $~$ $m_d^*$  $~$ & $~$ $m_u^*$  $~$ & $~$ $m_d^*$ $~$ & $~$ $m_u^*$  $~$ & $~$ $m_d^*$  $~$ & $~$ $m_u^*$  $~$ & $~$ $m_d^*$ $~$ \\
		\hline
		$~ 0 ~$ & ~$~ 0.2565 ~$~ & ~$~ 0.2564 ~$~ & ~$~ 0.2564 ~$~ & ~$~ 0.2564 ~$~ & ~$~ 0.2564 ~$~ & ~$~ 0.2564 ~$~ & ~$~ 0.2563 ~$~ & ~$~ 0.2563 ~$~ \\
		\hline
		$~ 0.25 ~$ & ~$~ 0.2315 ~$~ & ~$~ 0.2315 ~$~ & ~$~ 0.22323 ~$~ & ~$~ 0.2312 ~$~ & ~$~ 0.2351 ~$~ & ~$~ 0.2351 ~$~ & ~$~ 0.2357 ~$~ & ~$~ 0.2347 ~$~ \\
		\hline
		$~ 0.50 ~$ & ~$~ 0.2073 ~$~ & ~$~ 0.2073 ~$~ & ~$~ 0.2094 ~$~ & ~$~ 0.2070 ~$~ & ~$~ 0.2147 ~$~ & ~$~ 0.2147 ~$~ & ~$~ 0.2161 ~$~ & ~$~ 0.2141 ~$~ \\
		\hline
		$~ 0.75 ~$ & ~$~ 0.1841 ~$~ & ~$~ 0.1841 ~$~ & ~$~ 0.1877 ~$~ & ~$~ 0.1841 ~$~ & ~$~ 0.1953 ~$~ & ~$~ 0.1953 ~$~ & ~$~ 0.1975 ~$~ & ~$~ 0.1944 ~$~ \\
		\hline
		$~ 1 ~$ & ~$~ 0.1625 ~$~ & ~$~ 0.1625 ~$~ & ~$~ 0.1676 ~$~ & ~$~ 0.1627 ~$~ & ~$~ 0.1769 ~$~ & ~$~ 0.1769 ~$~ & ~$~ 0.1800 ~$~ & ~$~ 0.1758 ~$~ \\
		\hline
		$~ 2 ~$ & ~$~ 0.0990 ~$~ & ~$~ 0.0990 ~$~ & ~$~ 0.1091 ~$~ & ~$~ 0.1006 ~$~ & ~$~ 0.1183 ~$~ & ~$~ 0.1183 ~$~ & ~$~ 0.1242 ~$~ & ~$~ 0.1165 ~$~ \\
		\hline
		$~ 3 ~$ & ~$~ 0.0676 ~$~ & ~$~ 0.0676 ~$~ & ~$~ 0.0791 ~$~ & ~$~ 0.0699 ~$~ & ~$~ 0.0835 ~$~ & ~$~ 0.0835 ~$~ & ~$~ 0.0911 ~$~ & ~$~ 0.0820 ~$~ \\
		\hline
		$~ 4 ~$ & ~$~ 0.0518 ~$~ & ~$~ 0.0518 ~$~ & ~$~ 0.0630 ~$~ & ~$~ 0.0542 ~$~ & ~$~ 0.0638 ~$~ & ~$~ 0.0638 ~$~ & ~$~ 0.0722 ~$~ & ~$~ 0.00630 ~$~ \\
		\hline
		$~ 5 ~$ & ~$~ 0.0427 ~$~ & ~$~ 0.0427 ~$~ & ~$~ 0.0532 ~$~ & ~$~ 0.0451 ~$~ & ~$~ 0.0521 ~$~ & ~$~ 0.0521 ~$~ & ~$~ 0.0605 ~$~ & ~$~ 0.0518 ~$~ \\
		\hline
	\end{tabular}
	\caption{Observed effective quark masses corresponding to different values of baryonic densities $\rho_B$, isospin asymmetry $\eta$ and temperature $T$ }
	\label{TabDiracSigma} 
\end{table} 
 
\begin{figure*}
\centering
\begin{minipage}[c]{0.98\textwidth}
(a)\includegraphics[width=4.5cm]{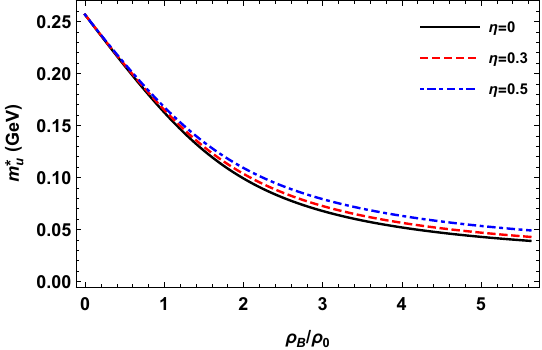}
\hspace{0.03cm}	
(b)\includegraphics[width=4.5cm]{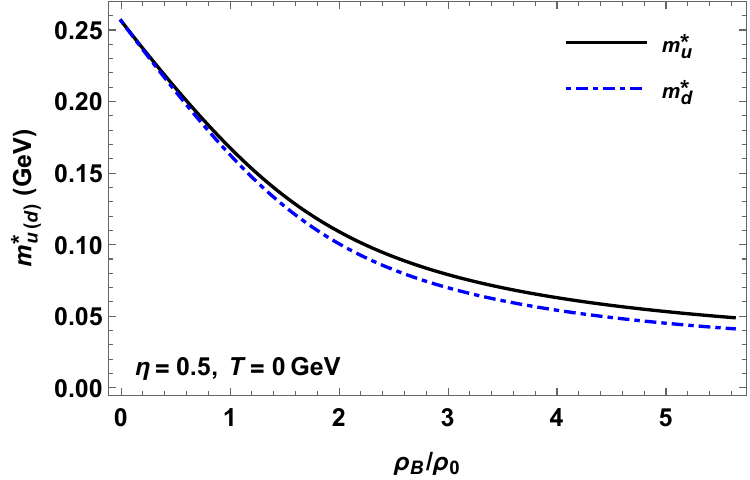}
\hspace{0.03cm}
(c)\includegraphics[width=4.5cm]{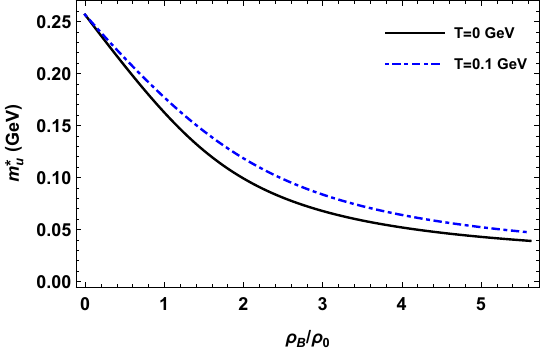}
\hspace{0.03cm}		
\end{minipage}
\caption{\label{fig1EffMass} (Color online) Effective mass for  $u$ quark   as a function of baryon density. Subplot (a) represents variation at temperature $T=0$ \, GeV for different values of asymmetry $\eta$, (b) 
 represents the
 effective quark masses $m_u^*$ and $m_d^*$ at zero temperature and $\eta = 0.5$,
and (c)
represents variation of effective mass for different values of temperature $T$ at $\eta = 0$.}
\end{figure*}

In Fig. \ref{fig1EffMass}, we have shown the variation of effective quark masses, $m_i^*$ of light quarks as a function of baryon density, $\rho_B/\rho_0$ (in units of nuclear saturation density $\rho_0$) of the nuclear medium. In Fig. \ref{fig1EffMass}(a), 
the effective quark mass $m_u^*$ has been plotted for isospin asymmetries $\eta = 0, 0.3$ and $0.5$. To have a quantitative understanding of this variation, the effective quark masses corresponding to some selective densities have been tabulated in Table \ref{TabDiracSigma}. A finite isospin asymmetry in the medium causes mass splitting between the members of isospin quark doublet $(u, d)$ leading to an increase in mass of $u$ and a drop for the $d$ quark as a function of $\eta$. This is true for a fixed value of density and temperature of the medium (also see Fig. \ref{fig1EffMass}(b) at $ \eta = 0.5$). Increase of the density of medium is observed to enhance the impact of isospin asymmetry of the medium. As the coupling strength of $s$ quark with the scalar-isovector field $\delta$ is zero ($g_\delta^s = 0$), the effective quark mass, $m_s^{*}$ is not affected considerably due to finite isospin asymmetry in the medium.
As shown in Fig. \ref{fig1EffMass}(c), the finite temperature of the medium is observed to cause an increase in the effective masses of different quarks. For the calculations of in-medium pion DAs and PDFs in LCQM, we have treated effective quark masses ($m^*_{u}$ and $m^*_{\bar{d}}$) and harmonic scale $\beta$ (fixed at $0.41$) as the input parameters. 

\begin{figure}
\centering
\begin{minipage}[c]{0.98\textwidth}
(a)\includegraphics[width=7.5cm]{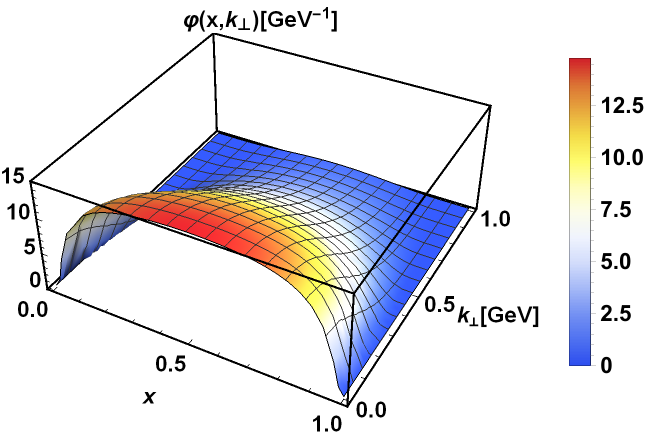}
\hspace{0.03cm}
(b)\includegraphics[width=7.5cm]{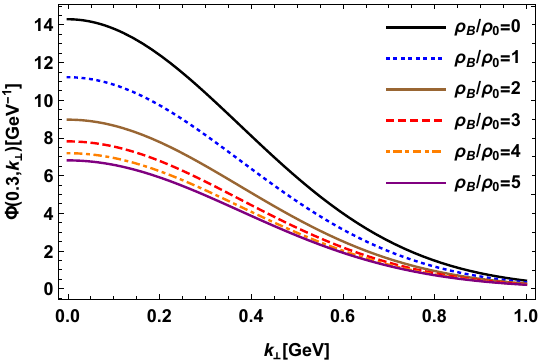}
\hspace{0.03cm}		
\end{minipage}
\caption{\label{fig2Wavefn} (Color online) (a) Dependence of light-cone momentum space wave function $\varphi (x,\bfk)$ on longitudinal momentum fraction $x$ and transverse momentum $\bfk$ in vacuum. (b) Comparison of in-medium light-cone wave functions with respect to $\bfk$ for different values of baryonic densities at fixed longitudinal momentum fraction $x=0.3$ for the case of $T=0$ and $\eta=0$.}
\end{figure}

\subsection{Pion Wave Function and Distribution Amplitudes}
In order to study the dependence of pion momentum space wave function on transverse momentum and longitudinal momentum fraction, in Fig. \ref{fig2Wavefn}(a) we have plotted the vacuum three dimensional structure of pion momentum space wave function as a function of transverse momentum $\bfk$ and longitudinal momentum fraction $x$. To analyze the impact of baryon density over the pion momentum space wave function, two dimensional plot with respect to transverse momenta $\bfk$ at fixed longitudinal momentum fraction $x=0.3$ for discrete values of baryon density $\rho_B$ has also been presented in Fig. \ref{fig2Wavefn}(b). The momentum space wave function shows a dramatic decrease in distribution at lower values of transverse momentum with increasing $\rho_B$.

At large momentum transfer, DAs can be analyzed through the exclusive processes. One can easily access the light-cone distributions via LCWFs by integrating out the transverse momentum. The correlation for defining the pseudo-scalar  DAs can be expressed as follows \cite{Li:2017mlw, Bodwin:2006dm,Choi:2007yu}
\be 
\langle 0|\bar{\vartheta}(z) \gamma^+ \gamma_5 \vartheta(-z)|\mathcal{P} (P^+,\bfP) \rangle = i k^+ \kappa_\pi \int_{0}^{1} dx \, e^{i(x-1/2) k^+ z^- } \phi (x) \bigg|_{z^+,\bfz=0} \, ,
\label{DA}
\ee
where $\vartheta$ represents the quark field operator and $\kappa_\pi$ is the decay constant. On substituting the meson state from Eq. (\ref{eqnq}) and quark field operators in Eq. (\ref{DA}), one can have the medium modified DAs $\phi^\ast (x,m^\ast_u, m^\ast_{\bar {d}})$ (henceforth to be referred as $\phi(x)$) in the form of LCWFs as 
\be 
\frac{\kappa^\ast_\pi}{2 \sqrt{2 N_c}} \phi (x)=\frac{1}{\sqrt{2x(1-x)}} \int \frac{d^2 \bfk}{16 \pi^3} [\psi (x,\bfk,\uparrow,\downarrow)-\psi (x,\bfk,\downarrow,\uparrow)] \, ,
\ee 
with $N_c=3$ being the number of colors of a quark flavor. The pion DA is normalized as 
\be 
\int_{0}^{1} dx \, \phi(x)=1 \, .
\ee

\begin{table}[h]
	\centering
	\begin{tabular}{|c|c|c|c|c|c|c|}
		\hline
		$~  ~$ & \multicolumn{6}{c|}{ ~$~$~ Medium modified decay constant ratio ($\kappa_\pi^\ast / \kappa_\pi$) ~$~$~} \\
		\cline{2-7}
		$~~ \rho_B/\rho_0 ~~$ & \multicolumn{4}{c|}{ $T=0$ GeV}  & \multicolumn{2}{c|} { $T=0.1$ GeV}    \\
		\cline{2-7}
		$~$  $~$ & $~$ $\eta=0$ \cite{Gifari:2024ssz} $~$ & $~$ $\eta=0  $ \cite{deMelo:2016uwj} $~$ & $~~~$ $\eta=0$  $~~~$ & $~$ $\eta=0.5$ $~$  & $~~~$ $\eta=0$ $~~~$ & $~$ $\eta=0.5$ $~$ \\
		\hline
		$~ 0 ~$ & ~$~ 1 ~$~ & ~$~ 1 ~$~ & ~$~ 1 ~$~ & ~$~ 1 ~$~ & ~$~ 1 ~$~ & ~$~ 1 ~$~ \\
		\hline
		$~ 0.25 ~$ & ~$~ 0.989 ~$~ &  ~$~ 0.868 ~$~ & ~$~ 0.953 ~$~ & ~$~ 0.953 ~$~ & ~$~ 0.959 ~$~ & ~$~ 0.960 ~$~  \\
		\hline
		$~ 0.50 ~$ &~$~ 0.978 ~$~ & ~$~ 0.730 ~$~ & ~$~ 0.898 ~$~ & ~$~ 0.902 ~$~ & ~$~ 0.915 ~$~ & ~$~ 0.917 ~$~    \\
		\hline
		$~ 0.75 ~$ & ~$~ 0.968 ~$~& ~$~ 0.592 ~$~ & ~$~ 0.84 ~$~ & ~$~ 0.848 ~$~ & ~$~ 0.868 ~$~ & ~$~ 0.873 ~$~   \\
		\hline
		$~ 1 ~$ &~$~ 0.957 ~$~ &  ~$~ 0.432 ~$~  & ~$~ 0.778 ~$~ & ~$~ 0.792 ~$~ & ~$~ 0.820 ~$~ & ~$~ 0.827 ~$~  \\
		\hline
		$~ 2 ~$ & ~$~ 0.903 ~$~ & ~$~ - ~$~ & ~$~ 0.555 ~$~ & ~$~ 0.593 ~$~ & ~$~ 0.631 ~$~ & ~$~ 0.650 ~$~  \\
		\hline
		$~ 3 ~$ & ~$~ - ~$~ &  ~$~ - ~$~ & ~$~ 0.414 ~$~ & ~$~ 0.463 ~$~ & ~$~ 0.487 ~$~ & ~$~ 0.518 ~$~  \\
		\hline
		$~ 4 ~$ & ~$~ - ~$~ & ~$~ - ~$~ & ~$~ 0.33 ~$~ & ~$~ 0.385 ~$~ & ~$~ 0.392 ~$~ & ~$~ 0.431 ~$~   \\
		\hline
		$~ 5 ~$ & ~$~ - ~$~ & ~$~ - ~$~ & ~$~ 0.279 ~$~ & ~$~ 0.334 ~$~ & ~$~ 0.331 ~$~ & ~$~ 0.372 ~$~   \\
		\hline
	\end{tabular}
	\caption{Medium modified decay constant ratio for pions corresponding to different values of baryonic density $\rho_{B}$, isospin asymmetry $\eta$ and temperature $T$. For comparison purpose, available data corresponding to $\eta=0$ and $T=0$ is tabulated from Ref. \cite{Gifari:2024ssz,deMelo:2016uwj}.}
	\label{TabDecayConst} 
\end{table} 
\begin{figure*}
\centering
\begin{minipage}[c]{0.98\textwidth}
(a)\includegraphics[width=7.5cm]{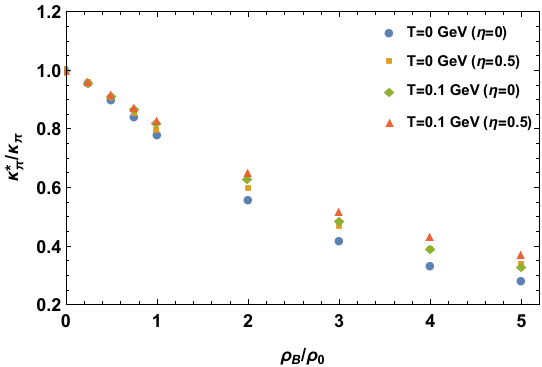}
\hspace{0.03cm}	
\end{minipage}
\caption{\label{figDecayConstant} (Color online) Decay constant ratios vs baryon density for different values of temperature $T$ and asymmetry $\eta$.}
\end{figure*} 
\begin{figure*}
\centering
\begin{minipage}[c]{0.98\textwidth}
(a)\includegraphics[width=7.5cm]{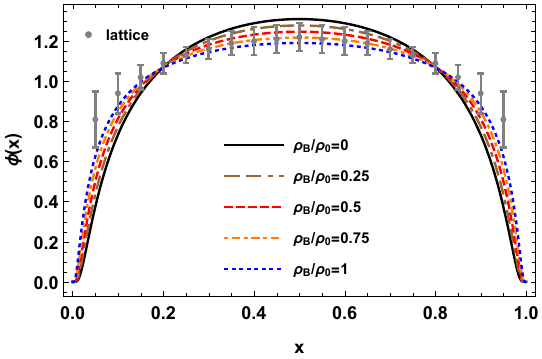}
\hspace{0.03cm}
(b)\includegraphics[width=7.5cm]{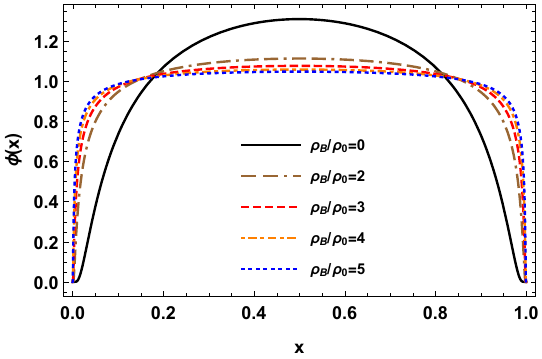}
\hspace{0.03cm}		
\end{minipage}
\caption{\label{fig3DAdensity} (Color online) Comparison of distribution amplitudes for valence $u$ quark of pion in vacuum and medium as a function of longitudinal momentum fraction in symmetric nuclear matter at zero temperature. Left panel represents the comparison of vacuum distribution with baryon density upto $\rho_0$ alongwith lattice simulation data \cite{LatticeParton:2022zqc} and right panel represents the comparison of vacuum distribution with baryon density above $\rho_0$ as well. }
\end{figure*} 

The medium effects on the decay constant ratio ${\kappa_\pi}^\ast$/ $\kappa_\pi$ have been calculated for pion at different densities in the present work and a comparison with other available model predictions \cite{Gifari:2024ssz,deMelo:2016uwj} have presented in Table. \ref{TabDecayConst}. In addition, a comparative behavior of the decay constant ratio with baryon density for different values of temperature and asymmetry have been presented in Fig. \ref{figDecayConstant}. In general, the decay constant ratio increases with increase of temperature and asymmetry. However, it has been found that at $T=0$ GeV, the decay constant ratio increases as the asymmetry of the matter increases and this increment is pronounced for baryonic density greater than $1$. At temperature $0.1$ GeV, the increment due to enhancement of asymmetry seems to be less as compare to decay constant ratio at $T=0$.  

\begin{figure*}
	\centering
	\begin{minipage}[c]{0.98\textwidth}
		(a)\includegraphics[width=7.5cm]{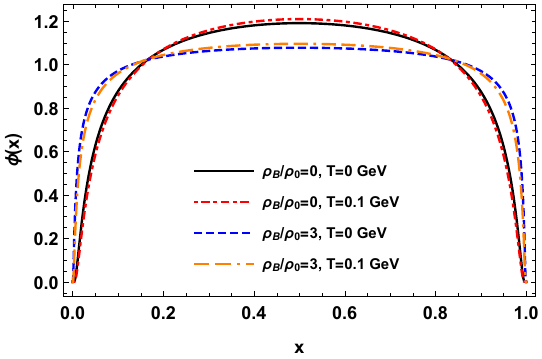}
		\hspace{0.03cm}
		(b)\includegraphics[width=7.5cm]{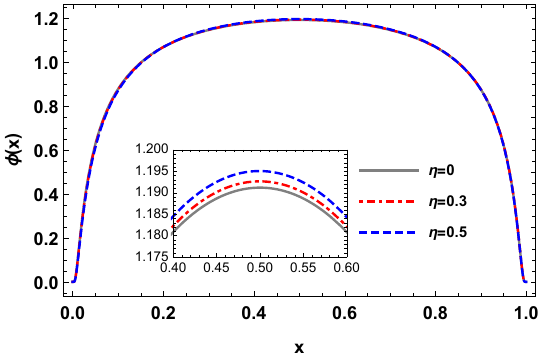}
		\hspace{0.03cm}		
	\end{minipage}
	\caption{\label{fig4DAtemp} (Color online) Comparison of vacuum and in-medium distribution amplitudes as a function of longitudinal momentum fraction for different values of temperature, $T$ [in subplot (a)]  and isospin asymmetry $\eta$ [in subplot (b)].}
\end{figure*}

We have demonstrated the vacuum and in-medium DAs in Fig. \ref{fig3DAdensity} for a wide range of baryon densities as a function of longitudinal momentum fraction $x$ and at a initial model scale $Q^2_{0}=0.23$ GeV$^2$. In Fig. \ref{fig3DAdensity}$(a)$, the dependency of distribution amplitude on densities from $\rho_{B} = 0$ to $\rho_0$ with a difference of $0.25$
is shown as a function of $x$. These results have  also been compared with the lattice data \cite{LatticeParton:2022zqc}.
Fig. \ref{fig3DAdensity}$(b)$ presents the results for DAs for higher baryon densities. We observe that an increase of baryon density from zero to a finite value causes   broadening of  distribution amplitude over the range of longitudinal momentum fraction. For a given value of $x$,  in the range $0.2<x<0.8$, the distribution amplitude decreases with increase  in the value of $\rho_B$. However, for longitudinal momentum fraction $x$, below 0.2 or above 0.8, value of $\phi(x)$ increases with  $\rho_B$.    This is primarily due to the scaling down of the in-medium quark masses $m_q^\ast$ which represents partial restoration of the chiral symmetry. Similar kind of in-medium dependence of DAs with baryon density has been observed in Refs. \cite{deMelo:2016uwj,Arifi:2023jfe}. In Ref. \cite{deMelo:2016uwj},  authors have  represented the in-medium DAs upto $\rho_{B} / \rho_{0}=1$ and have observed a dramatic decrease of DAs with increase in $\rho_B / \rho_{0}$ for the case of non normalized DAs. While the normalized DAs are consistent with our obtained results. It is clear from
 Fig. \ref{fig3DAdensity}$(b)$, for baryon densities above $2\rho_0$,   there is a significant shift in the amplitude  $\phi(x)$ to the smaller values with the flattening of distribution and broadening over almost complete region of longitudinal momentum fraction.
To understand the temperature dependence of DAs, in Fig. \ref{fig4DAtemp}(a) we have plotted pion DA at  baryon density $\rho_{B} =0$ and $3 \rho_0$ and have compared the results at temperature $T = 0$ and $0.1$ GeV. For the range $0.2<x<0.8$ of longitudinal momentum fraction,  $\phi(x)$ is   observed to increase with temperature $T$. However, beyond these values of $x$,    $\phi(x)$ decreases with increasing $T$.  This trend of variation of DA as a function of $T$ is opposite to the variation with   baryon density $\rho_B$ even though the impact of temperature is small as compared to that of baryon density. This is because of opposite impact of temperature and baryon density on the in-medium constituent quark masses. The effective mass $m_i^*$ increases with $T$  but decreases as a function of $\rho_B$.  As shown in Fig. \ref{fig4DAtemp}(b),   the impact of isospin asymmetry of the medium on $\phi(x)$   is observed to be minimal.

\begin{figure*}
	\centering
	\begin{minipage}[c]{0.98\textwidth}
		(a)\includegraphics[width=7.5cm]{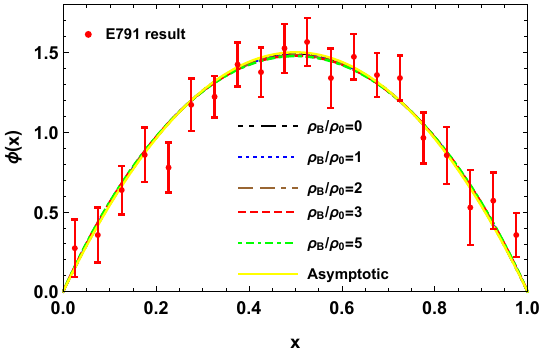}
		\hspace{0.03cm}
		(b)\includegraphics[width=7.5cm]{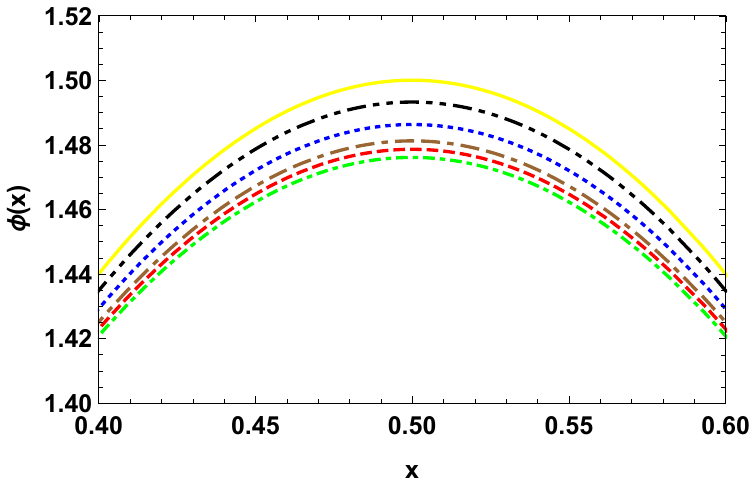}
		\hspace{0.03cm}		
	\end{minipage}
	\caption{\label{fig5DAexp} (Color online) Comparison of vacuum and in-medium distribution amplitudes (left panel) and magnified one (right panel) in symmetric nuclear matter at zero temperature, evolved from the model scale to $Q^2=10$ GeV$^2$ for valence $u$ quark of pion with experimental data, taken from Ref. \cite{E791:2000xcx}}
\end{figure*}

In order to study the behavior of DAs in high $Q^2$ region, results of the vacuum and in-medium evolved DAs have been represented in Fig. \ref{fig5DAexp} for baryon density ratios $\rho_B/\rho_0=0,1,2,3,5$ and a comparison with the experimental $E791$ data  has also been presented \cite{E791:2000xcx}. The leading order evolution for DAs has been carried out with Efremov-Radyushkin-Brodsky-Lepage ($ERBL$) equations \cite{Lepage:1980fj,Efremov:1979qk} to $Q^2=10$ GeV$^2$ from initial model scale $Q^2_0=0.23$ GeV$^2$ (Details have been given in Appendix A).
Compared to low $Q^2$ values, finite baryon density has less impact on the DA of pions at high $Q^2$. For better understanding of in-medium DAs, the magnified imaging of the distribution has been presented in Fig. \ref{fig5DAexp}(b) which clearly shows the decrement in the evolved DA distributions with an increase in $\rho_B$. This is for the range of $x$ shown in its subfigure. At $x = 0.5$, as the baryon density $\rho_B$ is increased from zero to $2\rho_0$,  the value of $\phi(x)$  decreases by $17.68$\%  and $1.4$\%, at $Q^2 = 0.23$ and $10$ GeV$^2$, respectively.   The evolved distributions represent a fair agreement with the experimental $E791$ data \cite{E791:2000xcx} and are also comparable to the asymptotic result of DA, $\phi_{asym}(x)=6x(1-x)$. Further details on the LO DA evolution have been provided in Sec. \ref{appendixa}.

\subsection{Parton Distribution Functions}
\begin{figure*}
\centering
\begin{minipage}[c]{0.98\textwidth}
(a)\includegraphics[width=7.5cm]{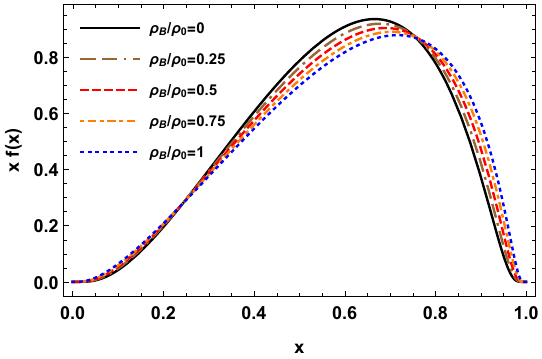}
\hspace{0.03cm}
(b)\includegraphics[width=7.5cm]{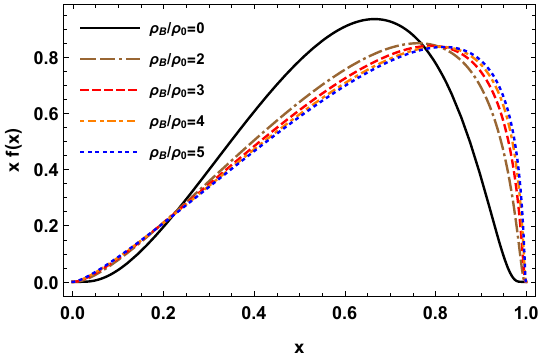}
\hspace{0.03cm}		
\end{minipage}
\caption{\label{fig6PDFs} (Color online) Comparison of vacuum and in-medium parton distribution functions for valence $u$ quark of pion  as a function of longitudinal momentum fraction in symmetric nuclear matter at zero temperature. Left panel represents the comparison of vacuum distribution with baryon density upto $\rho_0$ and right panel represents the comparison of vacuum distribution with baryon density above $\rho_0$.}
\end{figure*}

\begin{figure*}
\centering
\begin{minipage}[c]{0.98\textwidth}
\includegraphics[width=7.5cm]{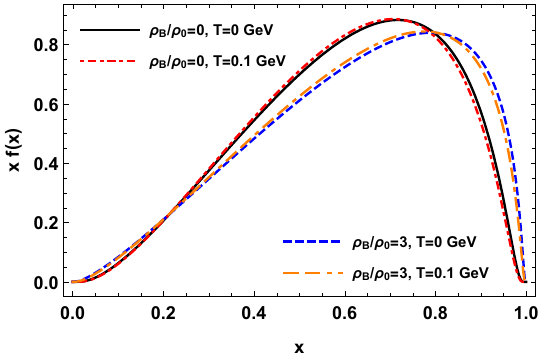}
\hspace{0.03cm}
\end{minipage}
\caption{\label{fig7PDFtemp} (Color online) Comparison of  parton distribution functions for valence $u$ quark of pion  as a function of longitudinal momentum fraction for different values of temperature and baryon density.}
\end{figure*}

The probability of finding the valence quark in pion with longitudinal momentum fraction $x$ can be accessed through the PDFs. At a fixed light-front time $\tau$, the correlator of the PDF is defined as \cite{Maji:2016yqo}
\be 
f(x)=\frac{1}{2} \int \frac{dz^-}{4 \pi} e^{i k^+ z_- /2} \langle \pi(P^+,\bfP;S)|\bar{\vartheta}(0) \Gamma \vartheta (z^-)|\pi(P^+,\bfP;S) \rangle |_{z^+, \bfz=0} \, ,
\ee
where spin $S=0$ for the pseudoscalar meson $\pi$. Using $\Gamma=\gamma^+$ in the above equation and  substituting the meson states from Eq. (\ref{eqnq}), the overlap form of unpolarized PDF $f(x,m_u^\ast,m_{\bar{d}}^\ast)$ (henceforth to be referred as $f(x)$) can be expressed as 
\be 
f(x)&=&\int \frac{d^2 \bfk}{16 \pi^3} \big[|\psi (x,\bfk,\uparrow,\uparrow)|^2 +|\psi (x,\bfk,\uparrow,\downarrow)|^2  + |\psi (x,\bfk,\downarrow,\uparrow)|^2 + |\psi (x,\bfk,\downarrow,\downarrow)|^2 \big] \, .
\ee
Its explicit form may be expressed as 
\be
f(x)&=& \int \frac{d^2 \bfk}{16 \pi^3} \bigg[\big((x {M}^*+m^*_{u})((1-x){M}^*+m^*_{\bar d})-\textbf{k}^2_\perp\big)^2
+\big({M}^*+ m^*_u+m^*_{\bar d}\big)^2\bigg] \frac{\mid \varphi^*(x,\textbf{k}_\perp)\mid^2}{\varUpsilon^{*2}_1 \varUpsilon^{*2}_2}. \label{pdf}
\ee
One can obtain the antiquark PDF of $\pi$ by using the $\bar f(1-x)$ function. However, due to equal mass of $u$ and $d$ quark in vacuum, both quark and antiquark PDFs are equal in vacuum and in-medium. Both vacuum and in-medium unpolarized PDF in Eq. (\ref{pdf}) obey the PDF sum rule \cite{Kaur:2020vkq,Puhan:2023ekt,Puhan:2023hio}
\be
\int d x f(x) =\int d x \bar f(1-x) =1,
\nonumber \\
\int d x [xf(x) + (1-x) f(x)] = 1.
\ee
Making use of the LFWFs mentioned in Eqs. (\ref{SpinWfns}) and (\ref{momspace}), distributions of PDFs have been portrayed in Fig. \ref{fig6PDFs} to analyze its density dependence. Fig. \ref{fig6PDFs}(a) presents the vacuum and in-medium variation of PDFs as a function of $x$ for densities upto $\rho_0$. Changing the baryon density from zero to $\rho_0$ shifts the position of peak of parton distribution function from $x = 0.665$ to 0.715.  Also, the  amplitude, i.e., the value of  $xf(x)$ decreases with increase in $\rho_B$ for the longitudinal momentum fraction in the range $0.3<x<0.7$. However, for the other values of $x$ shown in figure, $xf(x)$ increases with $\rho_B$. Fig. \ref{fig6PDFs}(b) shows that the position of peak  drifts toward the larger value of the longitudinal momentum fraction as $\rho_B$ is increased to much higher value, for example, at $\rho_B = 5\rho_0$ peak is observed  at  $x=0.820$. Hence, one can conclude that if a pion is immersed in asymmetric dense nuclear medium, then its constituent valence quark $u$ will carry more fraction of longitudinal momentum. The impact of temperature on the   valence quark PDF is shown in Fig. \ref{fig7PDFtemp}.  As was the case for DAs of $\pi$ mesons, the behavior of PDFs
as a function of temperature is also observed to be opposite  to that of the variation as a function of baryon density, $\rho_B$.
   \par

\begin{figure*}
\centering
\begin{minipage}[c]{0.98\textwidth}
(a)\includegraphics[width=7.5cm]{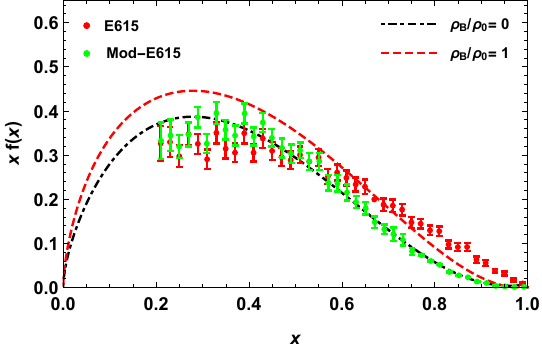}
\hspace{0.03cm}
(b)\includegraphics[width=7.5cm]{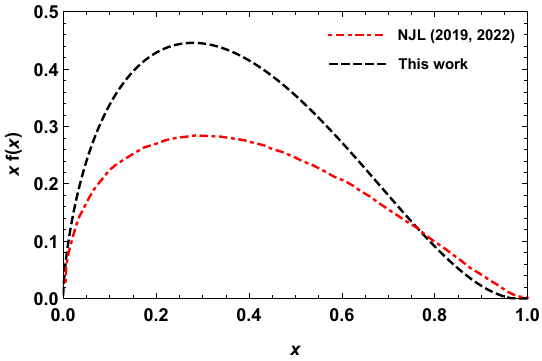}
\hspace{0.03cm}
\end{minipage}
\caption{\label{fig8EvolvedPDFs} (Color online) For valence $u$ quark of pion at zero temperature, (a) Comparison of vacuum and in-medium evolved parton distribution functions   with available experimental data from Refs.
\cite{Conway:1989fs,Aicher:2010cb} 
 and (b) Comparison of in-medium evolved parton distribution functions with NJL model results at baryon density $\rho_B = \rho_0$  \cite{Hutauruk:2019ipp,Hutauruk:2021kej}.}
\end{figure*} 
\begin{figure*}
\centering
\begin{minipage}[c]{0.98\textwidth}
\includegraphics[width=7.5cm]{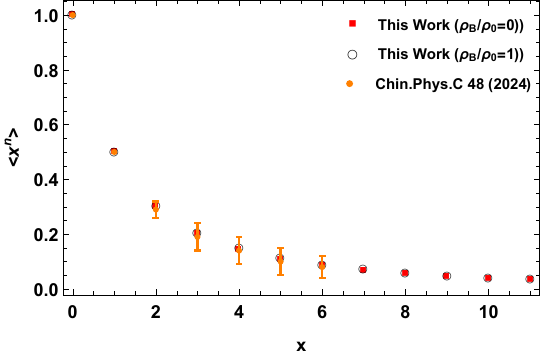}
\hspace{0.03cm}
\end{minipage}
\caption{\label{fig9} (Color online) Comparison of $u$-quark Mellin moments with Ref. \cite{Zhang:2023oja} for $\rho_B=0$ and $\rho_B=\rho_0$ at zero temperature.}
\end{figure*} 

The valence quark PDFs have been evolved using the next-to-leading order (NLO) Dokshitzer-Gribov-Lipatov-Altarelli-Parisi (DGLAP)
evolution equations \cite{Miyama:1995bd,Altarelli:1977zs} from our model scale to $16$ GeV$^2$ to compare our results with the available experimental results. We have compared our vacuum ($\rho_B=0$) and in-medium ($\rho_B= \rho_0$) evolved pion PDFs with E615 data \cite{Conway:1989fs} and modified E615 data \cite{Aicher:2010cb} in Fig. \ref{fig8EvolvedPDFs}(a). We observe that our vacuum result is in best agreement with modified E615 data. However, the evolved pion PDF at $\rho_B= \rho_0$ has higher amplitude of distribution than the vacuum and is in agreement at high $x$. The in-medium pion PDF is also found to be increasing with increase in $\rho_B$ at $Q^2=16$ GeV$^2$. However, in Refs. \cite{Hutauruk:2019ipp,Hutauruk:2021kej,Suzuki:1995vr}, the evolved in-medium pion PDF shows similar distributions as vacuum at high $Q^2=16$ GeV$^2$. We have compared our evolved in-medium PDF at $\rho_B= \rho_0$ with the available NJL model calculations \cite{Hutauruk:2019ipp,Hutauruk:2021kej} in Fig. \ref{fig8EvolvedPDFs} (b). In Ref. \cite{Hutauruk:2019ipp}, it has been observed that the kaon $u$-quark PDF shows increase in distribution with ian ncrease in $\rho_B$. Similar kind of observation seen in our calculations for pion $u$-quark PDF at high $Q^2$. Some details on the NLO running coupling constant have been provided in Sec. \ref{appendixb}.

Mellin moments of quark PDFs provide deep insights into the non-perturbative aspects of the QCD. Higher order Mellin moments give information about the quark densities with respect to different momentum fractions, whereas, the first moment correspond to the  average momentum fraction carried by the quark. The Mellin moments of unpolarized pion PDF can be calculated using \cite{Lu:2023yna}
\be
\langle x^n\rangle=\frac{\int dx ~ x^n f(x)}{\int dx ~ f(x)}.
\ee
In Fig. \ref{fig9}, we have presented the Mellin moments of $u$-quark PDF,  calculated at the model scale up to $n=11$,  for baryon density $\rho_B=0$ and $\rho_0$. The Mellin moments for in-medium PDFs have slightly higher values than the vacuum Mellin moment.
Our results of vacuum Mellin moment at model scale are found to be consistent with Ref. \cite{Zhang:2023oja}.

\section{Conclusion}
\label{Sec:conclusion}
In this work, we have presented the valence quark distributions for the lightest pseudoscalar meson i.e. pions  in the isospin asymmetric nuclear matter at zero and finite temperature. A hybrid approach, comprising of   light cone quark model and chiral SU($3$) quark mean field model, has been utilized for the  present study.  The distribution amplitudes and valence quark parton distribution function for pions have been calculated using the light-cone quark model and the required medium modified quark masses have been computed from the chiral SU($3$) quark mean field model. In the CQMF model,  the constituent quark masses in isospin asymmetric matter  have been modified through the scalar fields $\sigma, \zeta$ and $\delta$. For the present calculations, we have considered the medium with finite isospin asymmetry, i.e., different number of protons and neutrons.    The LCQM utilized for the calculations of pion properties has been introduced with dynamic spin effect. By solving the quark-quark correlation function for spin-0 mesons, with medium induced input parameters (medium modified effective quark masses), we have obtained the DAs and PDFs in the overlap form of light-cone wave functions in LCQM. 

We have studied the DAs for pion in vacuum as well as in-medium at different baryon density $\rho_B$, temperature $T$, and asymmetry $\eta$. Compared to vacuum case, the in-medium pion DA distributions shows broadening of peak over a wide range of longitudinal momentum fraction $x$. An increase of baryon density causes a decrease (for the range $0.2 <x<0.8$) whereas the temperature and isopsin asymmetry increases the DA value. The density effects are observed to be larger compared to temperature and isospin asymmetry. At model scale, both in-medium and vacuum DAs have been compared with lattice QCD simulations and are found to be in good agreement with lattice and other model predictions. DAs with different $\rho_B$ have been evolved to $Q^2 = 10$ GeV$^2$ using the ERBL equations and compared with experimental E791 data along with asymptotic result. We have observed that the evolved DAs distributions show very less difference for different $\rho_B$ than the model scale DAs. This indicates that the nuclear medium effect dominates in the non-perturbative region as compared to that in the perturbative region for DAs. However, our evolved DAs are found to be consistent with experimental data at high $Q^2$. 

The unpolarized $f(x)$ PDFs, obeying PDFs sum rule, have also been calculated for the valence $u$-quark of pion in LCQM. The constituent valence quark is observed to  carry more longitudinal momentum fraction pion in the dense nuclear medium as compared to that in the vacuum. We have evolved $x f(x)$ PDF at fixed $\rho_B=0$ and $\rho_B=\rho_0$ through NLO DGLAP evolutions from initial model scale to $16$ GeV$^2$ to compare our calculation with experimental $E615$ data as well as other model predictions. We have found that our evolved $x f(x)$ at $\rho_B=0$ almost matches with modified $E615$ experimental data, whereas it matches with a few $E615$ experimental data points at $\rho_B=\rho_0$. We have observed that the evolved PDF increases with an increase in $\rho_B$ indicating an increase in valence quark contribution to total PDFs. On the other hand, in case of vacuum, the valence quark contribution decrease at high $Q^2$.
We have also calculated the Mellin moment of $\langle x^n \rangle$ up to $n=11$  and compared with other model prediction at our model scale. In future, we aim to compute the impact of nuclear medium  on form factors and charge radii of pseudo-scalar mesons as well as DAs and PDFs for kaons along with the distribution function like TMDs and GPDs.

\section{Appendix A}\label{appendixa}
\textbf{Distribution Amplitude Evolution of Pion}

The leading order (LO) DAs evolution carried out by using the Efremov-Radyushkin-Brodsky-Lepage (ERBL) equations \cite{Lepage:1980fj,Efremov:1979qk}. In
a Gegenbauer basis \cite{RuizArriola:2002bp}, the evolved DAs can be expressed  as
\be
\phi(x,Q^2) =6x (1-x)\sum_{n=0}^\infty C_n^{\frac{3}{2}}(2x-1) a_n(Q^2),
\ee
with
\begin{eqnarray}
	a_n(Q^2)&=&\frac{2(2n+3)}{3(n+1)(n+2)}\bigg(\frac{\alpha_s(Q^2)}{\alpha_s(Q_{0}^2)}\bigg)^{\frac{\gamma^{(0)}_n}{2\beta_0}}\times \int_0^1 dx C_n^{\frac{3}{2}}(2x-1)\phi(x,Q_{0}^2),
\end{eqnarray}
where $C_n^{\frac{3}{2}}(2x-1)$ is a Gegenbauer polynomial.
The strong coupling constant $\alpha_s(Q^2)$ is given by
\begin{eqnarray}
	\alpha^{(LO)}_s(Q^2)=\frac{4\pi}{\beta_0 ~\ln\left(\frac{Q^2}{\Lambda^2_{QCD}}\right)}.
\end{eqnarray}
The factor $\frac{\gamma^{(0)}_n}{2\beta_0}$ defines the anomalous dimensions
\begin{eqnarray}
	\gamma^{(0)}_n=-2 c_F\bigg(3+\frac{2}{(n+1)(n+2)}-4 \sum_{m=1}^{n+1}\frac{1}{m}\bigg),
\end{eqnarray}
and
\begin{eqnarray}
	\beta_0=\frac{11}{3}c_A-\frac{2}{3}n_F,
\end{eqnarray}
where $c_A=3$ and $n_F$ correspond to the number of active flavors. The color factor $c_F=\frac{4}{3}$, initial model scale $Q_0^2=0.23$ GeV$^2$ and $\Lambda_{\mathrm{QCD}}=0.226$ GeV \cite{Gutsche:2014yea}.

\section{Appendix B}\label{appendixb}
\textbf{{PDFs evolutions}}

The quark PDFs evolution has been carried out using the next-to-leading order (NLO) Dokshitzer-Gribov-Lipatov-Altarelli-Parisi (DGLAP)
evolution equations \cite{Miyama:1995bd,Altarelli:1977zs}.  The NLO running coupling constant is given as
\begin{eqnarray}
	\alpha_s^{NLO}=\frac{4 \pi}{\beta_0 ln\left(\frac{Q^2}{\Lambda^2_{QCD}}\right)}\left(1- \frac{\beta_1 ln ln\left(\frac{Q^2}{\Lambda^2_{QCD}}\right)}{\beta^2_0 ln\left(\frac{Q^2}{\Lambda^2_{QCD}}\right)}\right).
\end{eqnarray}
The form $\beta_1$ is given by
\begin{eqnarray}
	\beta_1=\frac{34}{3}c^2_A -\frac{10}{3}c_A n_F -2 c_F n_F.
\end{eqnarray}


\begin{thebibliography}{200}
\section*{References} 
\bibitem{EuropeanMuon:1983wih}
J.~J.~Aubert, \textit{et al.} [European Muon],
Phys. Lett. B \textbf{123}, 275-278 (1983).

\bibitem{Suzuki:2002ae}
K.~Suzuki, \textit{et al.}
Phys. Rev. Lett. \textbf{92}, 072302 (2004).

\bibitem{Friedman:2004jh}
E.~Friedman, \textit{et al.}
Phys. Rev. Lett. \textbf{93}, 122302 (2004).

\bibitem{CHAOS:1996nql}
F.~Bonutti \textit{et al.} [CHAOS],
Phys. Rev. Lett. \textbf{77}, 603-606 (1996).

\bibitem{CHAOS:2004rhl}
P.~Camerini \textit{et al.} [CHAOS],
Nucl. Phys. A \textbf{735}, 89-110 (2004).


\bibitem{Fuchs:2004uu}
C.~Fuchs and A.~Faessler,
Prog. Part. Nucl. Phys. \textbf{53}, 59-75 (2004).


\bibitem{Kumano:2022cje}
S.~Kumano,
MDPI Physics \textbf{4}, 565-577 (2022).

\bibitem{AbdulKhalek:2021gbh}
R.~Abdul Khalek, \textit{et al.}
Nucl. Phys. A \textbf{1026}, 122447 (2022).


\bibitem{Hutauruk:2016sug}
P.~T.~P.~Hutauruk, I.~C.~Cloet and A.~W.~Thomas,
Phys. Rev. C \textbf{94}, 035201 (2016).

\bibitem{Nambu:1961tp}
Y.~Nambu and G.~Jona-Lasinio,
Phys. Rev. \textbf{122}, 345-358 (1961).

\bibitem{Puhan:2023ekt}
S.~Puhan, \textit{et al.}
JHEP \textbf{02}, 075 (2024)

\bibitem{Raha:2008ve}
U.~Raha and A.~Aste,
Phys. Rev. D \textbf{79}, 034015 (2009).

\bibitem{Nam:2007gf}
S.~i.~Nam and H.~C.~Kim,
Phys. Rev. D \textbf{77}, 094014 (2008).

\bibitem{Bijnens:2002hp}
J.~Bijnens and P.~Talavera,
JHEP \textbf{03}, 046 (2002)

\bibitem{daSilva:2012gf}
E.~O.~da Silva, \textit{et al.}
Phys. Rev. C \textbf{86}, 038202 (2012).

\bibitem{Koponen:2017fvm}
J.~Koponen, \textit{et al.}
Phys. Rev. D \textbf{96}, 054501 (2017).

\bibitem{Hedditch:2007ex}
J.~N.~Hedditch, \textit{et al.}
Phys. Rev. D \textbf{75}, 094504 (2007).
\bibitem{Suzuki:1995vr}
K.~Suzuki,
Phys. Lett. B \textbf{368}, 1-6 (1996).

\bibitem{deMelo:2014gea}
J.~P.~B.~C.~de Melo,  \textit{et al.}
Phys. Rev. C \textbf{90}, 035201 (2014).

\bibitem{Hutauruk:2018qku}
P.~T.~P.~Hutauruk, Y.~Oh and K.~Tsushima,
Phys. Rev. C \textbf{99},  015202 (2019).

\bibitem{Hutauruk:2019ipp}
P.~T.~P.~Hutauruk,  \textit{et al.}
Phys. Rev. D \textbf{100}, 094011 (2019).



\bibitem{Hutauruk:2021kej}
P.~T.~P.~Hutauruk and S.~i.~Nam,
Phys. Rev. D \textbf{105}, 034021 (2022).

\bibitem{Martin:1998sq}
A.~D.~Martin, \textit{et al.}
Eur. Phys. J. C \textbf{4}, 463, (1998).

\bibitem{Meissner:2008ay}
S.~Meissner, A.~Metz, M.~Schlegel and K.~Goeke,
JHEP \textbf{08}, 038 (2008).


\bibitem{Diehl:2015uka}
M.~Diehl,
Eur. Phys. J. A \textbf{52}, 149, (2016).

\bibitem{Angeles-Martinez:2015sea}
R.~Angeles-Martinez, \textit{et al.},
Acta Phys. Polon. B \textbf{46}, 2501, (2015).


\bibitem{Pasquini:2008ax}
B.~Pasquini, S.~Cazzaniga and S.~Boffi,
Phys. Rev. D \textbf{78}, 034025, (2008).

\bibitem{Diehl:2003ny}
M.~Diehl,
Phys. Rept. \textbf{388}, 41, (2003).

\bibitem{Kaur:2023zhn}
N.~Kaur and H.~Dahiya,
Eur. Phys. J. A \textbf{60}, 42(2024).

\bibitem{Serna:2020txe}
F.~E.~Serna, \textit{et al.}
Eur. Phys. J. C \textbf{80}, 955 (2020).

\bibitem{Chernyak:1983ej}
V.~L.~Chernyak and A.~R.~Zhitnitsky,
Phys. Rept. \textbf{112}, 173 (1984).




\bibitem{Brodsky:1997de}
S.~J.~Brodsky, H.~C.~Pauli and S.~S.~Pinsky,
Phys. Rept. \textbf{301}, 299-486 (1998).

\bibitem{Qian:2008px}
W.~Qian and B.~Q.~Ma,
Phys. Rev. D \textbf{78}, 074002 (2008).

\bibitem{Kaur:2019jow}
S.~Kaur and H.~Dahiya,
Phys. Rev. D \textbf{100}, 074008 (2019).

\bibitem{Wang:2001jw}
P.~Wang, Z.~Y.~Zhang and Y.~W.~Yu,
Commun. Theor. Phys. \textbf{36}, 71 (2001).

\bibitem{Kaur:2020vkq}
S.~Kaur,  \textit{et al.}
Phys. Rev. D \textbf{102}, 014021 (2020).

\bibitem{Xiao:2002iv}
B.~W.~Xiao, X.~Qian and B.~Q.~Ma,
Eur. Phys. J. A \textbf{15}, 523-527 (2002).

\bibitem{Ma:1993ht}
B.~Q.~Ma,
Z. Phys. A \textbf{345}, 321-325 (1993).

\bibitem{Brodsky:1989pv}
S.~J.~Brodsky and G.~P.~Lepage,
Adv. Ser. Direct. High Energy Phys. \textbf{5}, 93-240 (1989).

\bibitem{Lepage:1980fj}
G.~P.~Lepage and S.~J.~Brodsky,
Phys. Rev. D \textbf{22}, 2157 (1980).

\bibitem{Soper:1996sn}
D.~E.~Soper,
Nucl. Phys. B Proc. Suppl. \textbf{53}, 69-80 (1997).

\bibitem{Martin:2009iq}
A.~D.~Martin, \textit{et al.}
Eur. Phys. J. C \textbf{63}, 189-285 (2009).

\bibitem{Papazoglou:1998vr}
P.~Papazoglou, \textit{et al.}
Phys. Rev. C \textbf{59}, 411 (1999).
\bibitem{Wang:2001hw}
P.~Wang,  \textit{et al.}
Nucl. Phys. A \textbf{688}, 791 (2001).

\bibitem{Singh:2017mxj}
H.~Singh, A.~Kumar and H.~Dahiya,
Eur. Phys. J. Plus \textbf{134},   128 (2019).

\bibitem{Singh:2016hiw}
H.~Singh, A.~Kumar and H.~Dahiya,
Chin. Phys. C \textbf{41},  094104 (2017).



\bibitem{Singh:2018kwq}
H.~Singh, A.~Kumar and H.~Dahiya,
Eur. Phys. J. A \textbf{54},  120 (2018).

\bibitem{Singh:2020nwp}
H.~Singh, A.~Kumar and H.~Dahiya,
Eur. Phys. J. Plus \textbf{135},  422 (2020).

\bibitem{Kumar:2023owb}
A.~Kumar, S.~Dutt and H.~Dahiya,
Eur. Phys. J. A \textbf{60},  4 (2024).

\bibitem{Barik:1985rm}
N.~Barik, B.~K.~Dash and M.~Das,
Phys. Rev. D \textbf{31}, 1652 (1985).

\bibitem{Barik:2013lna}
N.~Barik, \textit{et al.}
Phys. Rev. C \textbf{88},   015206 (2013).


\bibitem{Huang:1994dy}
T.~Huang, B.~Q.~Ma and Q.~X.~Shen,
Phys. Rev. D \textbf{49}, 1490-1499 (1994).

\bibitem{Melosh:1974cu}
H.~J.~Melosh,
Phys. Rev. D \textbf{9}, 1095 (1974).

\bibitem{Xiao:2003wf}
B.~W.~Xiao and B.~Q.~Ma,
Phys. Rev. D \textbf{68}, 034020 (2003).

\bibitem{Yu:2007hp}
J.~h.~Yu, B.~W.~Xiao and B.~Q.~Ma,
J. Phys. G \textbf{34}, 1845-1860 (2007).


\bibitem{Arifi:2023jfe}
A.~J.~Arifi, P.~T.~P.~Hutauruk and K.~Tsushima,
Phys. Rev. D \textbf{107}, 114010 (2023).

\bibitem{deMelo:2016uwj}
J.~P.~B.~C.~de Melo, K.~Tsushima and I.~Ahmed,
Phys. Lett. B \textbf{766}, 125-131 (2017).

\bibitem{Mineo:2003vc}
H.~Mineo, \textit{et al.}
Nucl. Phys. A \textbf{735}, 482-514 (2004).




\bibitem{Li:2017mlw}
Y.~Li, P.~Maris and J.~P.~Vary,
Phys. Rev. D \textbf{96}, 016022 (2017).

\bibitem{Bodwin:2006dm}
G.~T.~Bodwin, D.~Kang and J.~Lee,
Phys. Rev. D \textbf{74}, 114028 (2006).

\bibitem{Choi:2007yu}
H.~M.~Choi and C.~R.~Ji,
Phys. Rev. D \textbf{75}, 034019 (2007).

\bibitem{Gifari:2024ssz}
G.~Gifari, P.~T.~P.~Hutauruk and T.~Mart,
[arXiv:2402.19048 [hep-ph]].

\bibitem{LatticeParton:2022zqc}
J.~Hua \textit{et al.} [Lattice Parton],
Phys. Rev. Lett. \textbf{129}, 132001 (2022).

\bibitem{E791:2000xcx}
E.~M.~Aitala \textit{et al.} [E791],
Phys. Rev. Lett. \textbf{86}, 4768-4772 (2001).



\bibitem{Efremov:1979qk}
A.~V.~Efremov and A.~V.~Radyushkin,
Phys. Lett. B \textbf{94}, 245-250 (1980).



\bibitem{Maji:2016yqo}
T.~Maji and D.~Chakrabarti,
Phys. Rev. D \textbf{94}, 094020 (2016).














\bibitem{Puhan:2023hio}
S.~Puhan and H.~Dahiya,
Phys. Rev. D \textbf{109}, 034005 (2024).







\bibitem{Miyama:1995bd}
M.~Miyama and S.~Kumano,
Comput. Phys. Commun. \textbf{94}, 185-215 (1996).

\bibitem{Altarelli:1977zs}
G.~Altarelli and G.~Parisi,
Nucl. Phys. B \textbf{126}, 298-318 (1977).

\bibitem{Conway:1989fs}
J.~S.~Conway, \textit{et al.}
Phys. Rev. D \textbf{39}, 92-122 (1989).





\bibitem{Aicher:2010cb}
M.~Aicher, A.~Schafer and W.~Vogelsang,
Phys. Rev. Lett. \textbf{105}, 252003 (2010).





\bibitem{Lu:2023yna}
Y.~Lu, \textit{et al.}
Phys. Lett. B \textbf{850}, 138534 (2024).




\bibitem{Zhang:2023oja}
S.~Zhang, X.~Wang, T.~Lin and L.~Chang,
Chin. Phys. C \textbf{48}, 033106 (2024).

\bibitem{RuizArriola:2002bp}
E.~Ruiz Arriola and W.~Broniowski,
Phys. Rev. D \textbf{66}, 094016 (2002).

\bibitem{Gutsche:2014yea}
T.~Gutsche, V.~E.~Lyubovitskij, I.~Schmidt and A.~Vega,
Phys. Rev. D \textbf{91}, 054028 (2015).








\end{thebibliography}
\end{document}